\documentclass{article}
\begin{document}

\title{QUANTUM EFFECTS IN THE SPACETIME OF A MAGNETIC FLUX COSMIC STRING}
\author{ M. E. X. GUIMAR\~AES$^1$ AND A. L. N. OLIVEIRA$^2$ \\
\mbox{\small{1. Departamento de Matem\'atica, Universidade de Bras\'{\i}lia}}\\
\mbox{\small{Campus Universit\'ario Darcy Ribeiro, CEP: 70910-900, Bras\'{\i}lia, Brazil}} \\
\mbox{\small{2. Instituto de F\'{\i}sica, Universidade de Bras\'{\i}lia}} \\
\mbox{\small{Campus Universit\'ario Darcy Ribeiro, CEP: 70910-900, Bras\'{\i}lia, Brazil}}}

\maketitle

\begin{abstract}
In this work we compute the vacuum expectation values of the energy-momentum 
 tensor and the average value of a massive, charged scalar field in the 
presence of a magnetic flux cosmic string for both zero- and 
finite-temperature cases.

\end{abstract}

\section{Introduction}  

In the General Relativity framework, a straight, infinite, static string lying
on the $z$-axis is described by a static metric with cylindrical symmetry~
\cite{linet}

\begin{equation}
\label{met}
ds^2 = -dt^2 + dz^2 + d\rho^2 + B^2 \rho^2 d\varphi^2
\end{equation}
where $\rho \geq 0$ and $0 \leq \varphi < 2\pi$ and the constant $B$ is related to the linear mass density $\mu$ as $B = 1-4\mu$.

Metric (\ref{met}) is locally but not globally flat.  The 
presence of the string leads to an azymuthal
deficit angle equal to $8\pi\mu$ and as a result this spacetime 
has a conical singularity. Placing the origin of the polar coordinate 
system on the string axis, one reveals a deficit $2\pi(1-B)$ of the polar 
angle $\varphi$.
Thus, near the string world sheet $\Sigma$ the space looks like the 
direct product $C_B \times \Sigma$ where $C_B$ is the conical space with the
corresponding ranging of the angle $0\leq \varphi\leq 2\pi B$.
Quantum effects such as the vacuum polarization effect arise in a 
flat spacetime whenever
the topology is non-trivial or boundaries are presented. A typical example 
is the Casimir effect~\cite{casimir} in which the nonzero expectation values 
of the energy-momentum tensor
owes its existence to the presence of boundaries in Minkowski spacetime. In 
the present case,
the manifold is complete, without boundaries, but topologically different 
from Minkowski space.

The phenomenom of vacuum polarization of a quantum field by a cosmic string 
carrying a internal magnetic flux can be understood as a realization in 
cosmology of the Aharonov-Bohm effect in electromagnetism~\cite{AB}. Indeed, 
a quantum field placed in the exterior of the string acquires an additional 
phase shift proportional to the magnetic flux  even though there is no 
magnetic field outside the string. Our purpose in this paper is to report 
and to show how to use some
results obtained in~\cite{mexg} concerning the vacuum polarization effect of 
a massive, charged scalar field in the spacetime (1) for zero- and 
finite-temperature cases. This problem has been treated by 
many authors~\cite{frolov}. The main contribution is to use a specific 
method for computing the Green's functions in spacetime (1) in which 
the renormalization procedure is straightforward.

Throughout this paper we adopt the system of units where $G=c=\hbar=1$.
\section{General Framework}

In order to study the vacuum polarization effect we need first to 
compute the scalar Green's function in spacetime (1). Once it is more 
convenient to work in the Euclidean approach to quantum theory we 
will in practice compute the Euclidean Green's function for the scalar field.
Therefore, let us consider the metric

\begin{equation}
\label{metric}
ds^2 = d\tau^2 +  dz^2 + d\rho^2 + B^2 \rho^2 d\varphi^2
\end{equation}
which is obtained from the metric (1) by a Wick rotation $t = -i\tau$ 
in the coordinate $t$. 

The Euclidean Green's function of a charged massive scalar 
field, $G_E^{(\gamma)}(x, x_0;m)$ is a solution of the covariant Laplace 
equation in the space (1)

\begin{equation}
\left[ \frac{\partial^2}{\partial\tau^2} + \frac{\partial^2}{\partial z^2}+ 
\frac{\partial^2}
{\partial\rho^2} + \frac{1}{\rho} \frac{\partial}{\partial\rho} + 
\frac{1}{B^2\rho^2}\frac{\partial^2}{\partial\varphi^2} - m^2 \right] 
G_E^{(\gamma)}(x, x_0;m) = - \frac{1}{B}\delta^{(4)}(x,x_0)
\end{equation}
and satisfies the following boundary conditions

\begin{eqnarray}
G_E^{(\gamma)}(\tau , z, \rho, \varphi + 2\pi) & = & 
e^{2i\pi\gamma}G_E^{(\gamma)}(\tau , z, \rho, \varphi)  \, , \\
\frac{\partial}{\partial \varphi}G_E^{(\gamma)}(\tau ,z, \rho, 
\varphi + 2\pi) & = & e^{2i\pi\gamma}\frac{\partial}{\partial \varphi}
G_E^{(\gamma)}(\tau ,z, \rho, \varphi )
\end{eqnarray}
where $\gamma$ is the fractional part of $\Phi/\Phi_0$, $\Phi_0$ being the 
flux quantum $2\pi/e$ and lies in the interval $ 0 \leq \gamma <1$. The case 
where $\gamma = 0$ corresponds 
to a vanishing flux whereas the case where $\gamma = 1/2$ corresponds to 
a twisted field around the axis $\rho = 0$. In addition to Eqs. (3-5), we 
also require that $G_E^{(\gamma)}(x, x_0;m)$ vanishes when $x$ and $x_0$ are 
infinitely separated. 

In order to solve Eq. (3), we will make a convenient change of coordinate 
$\theta = B \varphi$ in such a way that now we have to determine the Green's 
function in the subset of a Euclidean space covered by the coordinate system 
$(\tau, z, \rho, \theta)$ with $0 \leq \theta < 2\pi B$. 
Eq. (3) becomes the usual Laplace equation

\begin{equation}
\label{laplace}
(\Delta - m^2) G_E^{(\gamma)}(x, x_0;m) = - \delta ^{(4)}(x,x_0) \, .
\end{equation}
It is easy to see that the boundary conditions (4) and (5) will also modify 
under this new coordinate change. 

The important thing to notice here is that we can reduce our problem to 
2-dimensions by means of a  recurrence relation between the Green's 
functions of spaces of different dimensions

\begin{eqnarray}
G_{(\gamma)}^{N} (x^1, ..., x^{N-2}, \rho, \theta ;m) & = & \frac{1}{2} 
\int_{-\infty}^{\infty} G_{(\gamma)}^{N-1} (x^1, ..., x^{N-3}, \rho, \theta; 
\sqrt{m^2+ \lambda^2}) \nonumber \\
& &  \times\cos [\lambda (x^{N-2} - x_0^{N-2})] \, d\lambda \, .
\end{eqnarray}

After setting the relevant preliminary steps, we are now in a position 
to compute the Green's function for a massive scalar field.  We start by 
noting that the solution of Eq. (\ref{laplace}) is the usual Green's function

\begin{equation}
\frac{1}{2\pi} K_0(mr_2) \, ,
\end{equation}
where $r_2$ is the Euclidean distance between the two points $(\rho, \theta)$ 
and $(\rho_0, \theta_0)$ and $K_{\mu}$ denotes the modified Bessel function 
of second kind. However, clearly, 
this solution does not satisfy the boundary conditions (4) and (5). To solve 
this problem we propose a 
generalisation of the integral expression of the Bessel function in the form

\begin{equation}
\label{green}
G^{(2)}_{(\gamma)} = \frac{1}{2\pi^2} \int_{-\infty}^{\infty} 
K_{i\nu}(m\rho)K_{i\nu}(m\rho_0)
e^{\pi\nu}U_{\nu}(\theta) d\nu \, ,
\end{equation}
where $U_{\nu}$ must satisfy the boundary conditions (4) and (5). Therefore, 
we find that

\begin{equation}
\label{u}
U_{\nu}(\theta) = \frac{e^{2i\pi\gamma}\sinh(\nu\mid\theta -\theta_0\mid) - \sinh(\nu\mid\theta - \theta_0\mid - 2\pi\nu B)}{\cosh(2\pi\nu B) - \cos (2\pi\gamma)} \, .
\end{equation}

In this way, the Green's function (\ref{green}) with (\ref{u}) is a solution of the Laplace equation (\ref{laplace}) and satisfies the boundary conditions (4) and (5). In principle, our problem is solved and it would be enough to apply twice the recurrenc
e relation. However, we can still improve the expression of the Green's function by manipulating with its integral expressions. By doing this (for detail of this calculations, we refer the reader to Refs. 4) and by applying successively the recurrence rel
ation, we get a local form

\begin{equation}
\label{g}
G^{(4)}_{(\gamma)}(x,x_0;m) = \frac{mK_1(mr_4)}{4\pi^2 r_4} + G^{*(4)}_{(\gamma)}(x,x_0;m)
\end{equation}
valid in the domain $ \pi/B - 2\pi  < \varphi - \varphi_0 < 2\pi  - \pi/B $ for $B > 1/2$. The term $G^{*(4)}_{(\gamma)}(x,x_0;m)$ appearing in Eq. (\ref{g}) is a regular term which is given by 

\begin{equation}
G^{*(4)}_{(\gamma)}(x,x_0;m)= \frac{m}{8\pi^3 B} \int_0^{\infty} 
\frac{K_1[mR_4(u)]}{R_4(u)}F^{(\gamma)}_B(u, \varphi - \varphi_0) du \, ,
\end{equation}
with $R_4(u) = \sqrt{(\tau - \tau_0)^2 + (z - z_0)^2 + \rho^2 + \rho_0^2 + 2\rho\rho_0\cosh u}$ and the function \newline
$F^{(\gamma)}_B(u, \varphi - \varphi_0)$ contains the contributions coming from the Aharonov-Bohm and the non-trivial gravitational interactions between the scalar field and the cosmic string

\begin{eqnarray}
F^{(\gamma)}_B(u, \psi) &  = & i \frac{e^{i(\psi +\pi/B)\gamma}\cosh[u(1-\gamma)/B] - e^{-i(\psi + \pi/B)(1-\gamma)}\cosh[u\gamma/B]}{\cosh(u/B) - \cos(\psi + \pi/B)} \nonumber \\
& & - i \frac{e^{(\psi - \pi/B)\gamma}\cosh[u(1-\gamma)/B] - e^{-i(\psi - \pi/B)(1-\gamma)}\cosh[u\gamma/B]}{\cosh(u/B) - \cos(\psi -\pi/B)}
\end{eqnarray}
with $\psi \equiv \varphi - \varphi_0$. In the next section we will compute the vacuum expectation values of some physical quantities using the Euclidean Green's function given by Eq. (\ref{g}).

\section{Vacuum Polarization Effect}

The renormalization of the vacuum expectation values (V.E.V.) of some physical quantities is performed in a straightforward way and consists solely in removing the usual Green's function from Eq. (\ref{g}). In this section, we will compute the V.E.V. of t
he average value and the energy-momentum tensor of the scalar field. We will treat both the zero- and finite-temperature cases.

Let us suppose that the scalar field is in thermal equilibrium with finite temperature $T$. In this case, its (thermal) Green's function is a solution of Eq. (\ref{laplace}), satisfies boundary conditions (4) and (5) and, in addition, is periodic in the c
oordinate $\tau$ with a period equal to $\beta = 1/k T$, where $k$ is the Boltzmann constant. 

Metric (\ref{metric}) is ultrastatic (static and $g_{00}=1$). Therefore, we can apply de Schwinger-deWitt formalism~\cite{witt} 
\begin{equation}
\label{dewitt}
G_{ET}^{(\gamma)}(x,x_0;m) = \int_{0}^{\infty} K_{ET}^{(\gamma)}(x,x_0;s)ds
\end{equation}
and we can derive the Euclidean thermal heat kernel $K_{ET}(x, x_0;s)$ from its corresponding Euclidean zero-temperature heat kernel $K_E(x,x_0;s)$  becomes~\cite{page}
\begin{equation}
K_{ET}^{(\gamma)}(x,x_0;s) = \Theta_3 \left(i\frac{\beta(\tau - \tau_0)}{4s}\mid i\frac{\beta^2}{4\pi s} \right) K_E^{(\gamma)}(x, x_0;s) \, ,
\end{equation}
where $ \Theta_3 $ is defined as in Ref. 6. To simplify our problem, we will compute the Euclidean thermal Green's function for a massless scalar field $D_{ET}^{(\gamma)}(x,x_0)$. Also, we will skip here any details of the calculations and we will give th
e final results directly. Therefore, we have
\begin{eqnarray}
\label{thermal}
D_{ET}^{(\gamma)}(x,x_0) & = & \frac{1}{4\pi\beta d}
\frac{\sinh(2\pi/\beta)d}{[\cosh(2\pi/\beta)d - \cos(2\pi/\beta)(\tau - \tau_0)]} \nonumber \\
& & + \frac{1}{8\pi^2 B \beta}\int_0^{\infty} \frac{\sinh[(2\pi/\beta)D(u)]F_B^{(\gamma)}(u,\psi) du}{D(u)[\cosh(2\pi/\beta)D(u)- \cos(2\pi/\beta) (\tau - \tau_0)]} \, ,
\end{eqnarray}
where $d= \frac{\pi}{B}[(z-z_0)^2 + \rho^2 + \rho_0^2 - 2\rho\rho_0\cos B \psi]^{1/2}$ and $D(u) = \frac{\pi}{B}[(z-z_0)^2 + \rho^2\rho_0^2 + 
2\rho\rho_0^2 \cosh u]^{1/2}$. 

Using (\ref{thermal}), we can compute the (thermal) average $<\phi^2(x)>_{\beta}$ and the (thermal) V.E.V. of the energy-momentum tensor for a massless scalar field.  We can give only analytic results in the assymptotic limits. Therefore, in the limit $\b
eta \rightarrow \infty$ (or, equivalently $T \rightarrow 0$), we have:
\begin{equation}
<\phi^2(x)>_{\beta\rightarrow \infty} \sim \frac{\omega_2 (\gamma)}{2 \rho^2} \, ,
\end{equation}
\begin{eqnarray}
<T_{\mu\nu}>_{\beta\rightarrow \infty} & = & [\omega_4(\gamma) - \frac{1}{3}
\omega_2(\gamma)]\frac{1}{\rho^4} {\rm diag}(1,1,1,-3) \nonumber \\
& & + 2(\xi - \frac{1}{6})\omega_2(\gamma)\frac{1}{\rho^4} {\rm diag} (1,1,-\frac{1}{2},\frac{3}{2}) \, .
\end{eqnarray}
In the limit $\beta \rightarrow 0$, we have:
\begin{equation}
<\phi^2(x)>_{\beta\rightarrow 0} \sim \frac{1}{12\beta^2} + \frac{M^{(\gamma)}}{\beta\rho} \, ,
\end{equation}
\begin{eqnarray}
 <T^t_t>_{\beta\rightarrow 0} & \sim & -\frac{\pi^2}{15\beta^4} + (2\xi - 1/2)\frac{M^{(\gamma)}}{\beta\rho^3}  \, , \\
<T^z_z>_{\beta\rightarrow 0} & \sim &  -\frac{\pi^2}{45\beta^4} + \frac{N^{(\gamma)}}{\beta\rho^3} + (2\xi - 1/2)\frac{M^{(\gamma)}}{\beta\rho^3} \, , \\
<T^{\rho}_{\rho}>_{\beta\rightarrow 0} & \sim &  -\frac{\pi^2}{45\beta^4} + \frac{N^{(\gamma)}}{\beta\rho^3} - 2\xi\frac{M^{(\gamma)}}{\beta\rho^3} \, , \\
<T^{\varphi}_{\varphi}>_{\beta\rightarrow 0} & \sim &  -\frac{\pi^2}{45\beta^4} -2 \frac{N^{(\gamma)}}{\beta\rho^3} + 4\xi\frac{M^{(\gamma)}}{\beta\rho^3} \, ,
\end{eqnarray}
with $\omega_2(\gamma) \, , \omega_4(\gamma)$, evaluated numerically, and $M^{(\gamma)} \, N^{(\gamma)}$, given in integral forms, respectively:

\begin{eqnarray}
\omega_2(\gamma) & = & -\frac{1}{8\pi^2}\{ \frac{1}{3} - \frac{1}{2B^2}[ 4(\gamma - \frac{1}{2}^2 -\frac{1}{3} ]\} \, , \\
\omega_4(\gamma) & = & -\frac{1}{720\pi^2} \{ 11 -\frac{15}{B^4}[ 
4(\gamma -\frac{1}{2})^2 -\frac{1}{3}]  \, ,\nonumber \\
& & + \frac{15}{8B^4}[ 16(\gamma -\frac{1}{4})^4 - 8(\gamma -\frac{1}{2})^2 
+\frac{7}{15}] \}  \\
M^{(\gamma)} & = & \frac{1}{16\pi^2B} \int_0^{\infty} \frac{F_B^{(\gamma)}(u,0)}
{\cosh u/2} \, du \, , \\
N^{(\gamma)} & = & \frac{1}{32\pi^2B}\int_0^{\infty} \frac{F_B^{(\gamma)}(u,0)}
{\cosh^{3/2} u/2} \, du  \, .
\end{eqnarray}

\section{Concluding Remarks}

This work considered the vacuum polarization effect of a scalar field in 
the spacetime generated by a cosmic string. Due to the particular properties of 

this spacetime we could apply a method which allows one to write the Green's functions in a local form as a sum of the usual Green's function in Minkowski spacetime and a regular term which is responsible for the vacuum polarization effect. Although here 
we considered the metric of a cosmic string in General Relativity, we point out that this method can be extended to the case of a scalar-tensor theory of gravity as well.~\cite{bezerra}

\medskip
\noindent{\bf Acknowledgements} 
\medskip

One of the authors (A.L.N.O.) would like to thank CAPES for a PhD grant. This work was partially supported by FINATEC/UnB.

\end{document}